\documentstyle[aps,preprint,eqsecnum,floats,epsf]{revtex}
\tighten
\draft
\begin{document}

\title{Constraints on Neutrino Mixing}

\author{A.~B. Balantekin\thanks{Electronic address:
        {\tt baha@nucth.physics.wisc.edu}}}
\address{Department of Physics, University of Wisconsin\\
         Madison, Wisconsin 53706 USA\thanks{Permanent Address}\\
 and\\
         Institute for Nuclear Theory, University of Washington, Box
         351550\\
         Seattle, WA 98195-1550 USA}
\author{G.~M. Fuller\thanks{Electronic address:
        {\tt gfuller@ucsd.edu  }}}
\address{Department of Physics, University of California, San Diego\\
         La Jolla, CA 92093-0319 USA\thanks{Permanent Address}\\
 and\\
         Institute for Nuclear Theory, University of Washington, Box
         351550\\
         Seattle, WA 98195-1550 USA}

\maketitle

\begin{abstract}
  We explore the implications of imposing the constraint that two
  neutrino flavors (which for definiteness we take to be $\nu_{\mu}$  
and
$\nu_{\tau}$) are similarly coupled to the mass basis in addition to
  the unitarity constraints. We allow three active and an arbitrary
  number of sterile neutrinos. We show that in this scheme one of the
  mass eigenstates decouples from the problem, reducing the dimension  
of
  the flavor space by one.

\end{abstract}


%%%%%%%%%%%%%%%%%%%%%%%%%%%%%%%%%%%%%%%%%%%%%%%%%%%%%%%%%%%%%%%%%%%%%
%%%%%%%%%%%%%%%%%%%%%%%%%%%%%%%%%%%%%%%%%%%%%%%%%%%%%%%%%%%%%%%%%%%%%

\newpage

\section{Introduction}

\indent

Recent observations of atmospheric neutrinos at the SuperKamiokande
experiment \cite{atm} very strongly suggest that the muon neutrino
maximally mixes with another neutrino, which is either the tau
neutrino or a sterile neutrino. In a parallel development Elwood,
Irges, and Ramond suggested that the observed quark hierarchies
indicate a simple family symmetry \cite{ramond}. They argue that the
manifest interfamily hierarchy, for example in the Wolfenstein
parameterization \cite{lincoln} of the CKM matrix, points to the
existence of an anomalous $U(1)$ family symmetry beyond the Standard
Model. They conclude that the MNS neutrino mixing matrix \cite{mns}
will be of the form
\begin{equation}
\left(\matrix{
1 & \lambda^3   & \lambda^3   \cr
\lambda^3   & 1 & 1 \cr
\lambda^3   & 1 & 1 \cr }
\right)~~~ ,
\label{1}
\end{equation}
where $\lambda$ is the Cabibbo angle. Motivated by this work and the
Superkamiokande results in this paper we wish to explore the
implications of the constraint that two neutrino flavors (which for
illustrative purposes we
take to be $\nu_{\mu}$ and $\nu_{\tau}$) are similarly
coupled to the mass basis.

The possibility of the existence of light sterile neutrinos was
suggested not only by the atmospheric neutrino measurements, but by
the LSND experiment \cite{lsnd} and by the recent work
\cite{bf2m,cfq}
on the possibility of an active-sterile neutrino transformation
enabling the production of the $r$-process nuclei in neutrino-heated
supernova ejecta \cite{review}.  Hence we consider three active
flavors and an 
arbitrary number (which could be taken to be zero) of sterile
neutrinos. At this point we do not specify neutrino mass eigenvalues
and allow all possible schemes.  The $N \times N$ neutrino mixing
matrix will be denoted by $U_{\alpha i}$ where $\alpha$ denotes the
flavor index and $i$ denotes the mass index:
\begin{equation}
  \label{eq:2}
  | \nu_{\alpha} \rangle = \sum_i U_{\alpha i} | \nu_i \rangle.
\end{equation}
Note that even though we want to impose $|U_{\mu i}| \sim |U_{\tau i}|$
we cannot simply take $U_{\mu i} = U_{\tau i}$ as that gives $\det U
=0$, in contradiction with unitarity which requires $\det U =1$.
Instead we impose the next simplest constraint, namely that they are
proportional for all but one mass eigenstate, which we choose for
definiteness to be the third mass eigenstate:
\begin{equation}
  \label{eq:3}
  U_{\mu i} \sim  U_{\tau i} \neq 0, \> \forall i \neq 3.
\end{equation}
We write this condition in terms of an arbitrary angle $\phi$ and an
arbitrary phase $\eta$:
\begin{equation}
  \label{eq:15}
  \sin \phi \> U_{\mu i} =  e^{i\eta} \cos \phi \>
  U_{\tau i} \neq 0, \> \forall i \neq 3.
\end{equation}
Note that, in our formalism, we permit CP-violating phases.
Introducing the quantity
\begin{equation}
  \label{a1}
  A = \sum_{i \neq 3} \left[ |U_{\mu i}|^2 + |U_{\tau i}|^2 \right]
\end{equation}
and using Eq. (\ref{eq:15}) one can easily show that
\begin{equation}
  \label{a2}
  \sum_{i \neq 3} |U_{\mu i}|^2 = A \cos^2  \phi,
\end{equation}
and
\begin{equation}
  \label{a3}
  \sum_{i \neq 3} |U_{\tau i}|^2 = A \sin^2  \phi.
\end{equation}
Eqs. (\ref{a2}) and (\ref{a3}) along with the unitarity
constraints $\sum_i |U_{\mu i}|^2 = 1 = \sum_i |U_{\tau i}|^2$
immediately yield
\begin{equation}
  \label{eq:4}
   |U_{\mu 3}|^2 = 1 - A \cos^2  \phi,
\end{equation}
and
\begin{equation}
  \label{eq:4p}
   |U_{\tau 3}|^2 = 1 - A \sin^2  \phi.
\end{equation}
Inserting Eqs. (\ref{eq:4}) and (\ref{eq:4p}) into the unitarity
constraint $ \sum_i U_{\mu i} U_{\tau i}^{\ast} = 0$ and using the
constraint $\sum_i |U_{\mu i}|^2 = 1$ we obtain
\begin{equation}
  \label{a4}
  A =1,
\end{equation}
\begin{equation}
  \label{a5}
  U_{\mu 3} = - \sin \phi e^{i\delta} e^{i\eta},
\end{equation}
and
\begin{equation}
  \label{a6}
   U_{\tau 3} = \cos \phi  e^{i\delta},
\end{equation}
where $\delta$ is a phase to be determined.
Furthermore one can write down the unitarity constraints $\sum_i
U_{\mu i}U_{\alpha i}^{\ast} = 0$ and $\sum_i U_{\tau i}U_{\alpha
  i}^{\ast} = 0$, where $\alpha$ stands for either the electron
neutrino or one of the sterile neutrinos. Inserting
Eqs. (\ref{eq:15}), (\ref{a5}), and (\ref{a6})
into these constraints and subtracting them gives
\begin{equation}
  \label{eq:6}
  U_{\alpha 3} = 0 , \>\> \alpha \neq \mu,\tau,
\end{equation}
which implies
\begin{equation}
  \label{eq:6a}
  | \nu_{\alpha} \rangle = \sum_{i \neq 3} U_{\alpha i}
  | \nu_i \rangle, \>\> \alpha \neq \mu,\tau.
\end{equation}
This is a remarkable result. Simply imposing the restriction in Eq.
(\ref{eq:3}) and taking into account the unitarity of the neutrino
mixing matrix decouples all the other flavors from the third mass
eigenstate. Similarly introducing the states
\begin{equation}
  \label{eq:7}
  |  \tilde{\nu}_{\mu} \rangle =  \cos \phi
| \nu_{\mu} \rangle +  \sin \phi e^{i\eta} | \nu_{\tau} \rangle ,
\end{equation}
and
\begin{equation}
  \label{eq:8}
  |  \tilde{\nu}_{\tau} \rangle =  - \sin \phi e^{-i\eta} |
    \nu_{\mu} \rangle + \cos \phi | \nu_{\tau} \rangle  ,
\end{equation}
one can easily show that
\begin{equation}
  \label{eq:9}
  | \tilde{\nu}_{\mu} \rangle = \frac{1}{\cos \phi} \sum_{i \neq 3}
  U_{\mu i} | \nu_i \rangle,
\end{equation}
and
\begin{equation}
  \label{eq:10}
  | \tilde{\nu}_{\tau} \rangle =  e^{i \delta} | \nu_3 \rangle.
\end{equation}
The combination of flavor states in Eq. (\ref{eq:8}) is a pure mass
eigenstate and is decoupled from all the physical processes except  
the
$\nu_{\mu} \rightleftharpoons \nu_{\tau}$ oscillations (which could  
be
taken to fit the atmospheric neutrino results). The mixing of
the remaining neutrinos is governed by Eqs. (\ref{eq:6a}) and
(\ref{eq:9}), i.e. the original $N$ flavor mixing problem is reduced
to a problem where $N-1$ flavors (the electron neutrino,
$\tilde{\nu}_{\mu}$, and the sterile neutrinos) are mixed.

\section{Mixing of three active flavors}

If we have only three active flavors of neutrinos mixing with no
contribution from the sterile neutrinos the state $|
\tilde{\nu}_{\tau} \rangle$ decouples and the two states
\begin{equation}
  \label{eq:11}
  \left(\matrix{
      | \nu_e \rangle \cr
      | \tilde{\nu}_{\mu} \rangle  }
\right)~~~
\end{equation}
result from transforming the mass eigenstates with the matrix
\begin{equation}
  \label{eq:12}
  \left(\matrix{
     U_{e 1} & U_{e 2}  \cr
     U_{\mu 1}/ \cos \phi &  U_{\mu 2}/ \cos \phi  }
\right)~~.
\end{equation}
The solar neutrino data in this case could be explained by either the
matter-enhanced or vacuum $\nu_e \rightarrow \tilde{\nu}_{\mu}$
oscillation. Since both $\nu_{\mu}$ and $\nu_{\tau}$ gain the same
effective mass due to coherent forward scattering in matter, standard
MSW analyses for $\nu_e \rightarrow \nu_{\mu}$ conversion \cite{bks}
is also valid for $\nu_e \rightarrow \tilde{\nu}_{\mu}$ conversion.
For the solar neutrino experiments where mu and tau neutrinos are
detected (such as SuperKamiokande and SNO) it is not possible to
distinguish $\nu_{\mu}$ and $\tilde{\nu}_{\mu}$ as at the low solar
energies both of these neutrinos have the same neutral current
interactions with electrons and deuterons. The recoil electron  
kinetic
energy spectra for $\nu_e$ and $\tilde{\nu}_{\mu}$ are however
different. This difference can be used to determine the solar
$\tilde{\nu}_{\mu}$ component at BOREXINO or KamLAND \cite{muraya}.

In the special case $\phi = \pi/4$
the mixing matrix for the three active flavors is
\begin{equation}
  \label{eq:13}
  \left(\matrix{
     U_{e 1} & U_{e 2} & 0  \cr
     \sqrt{2} U_{\mu 1} & \sqrt{2} U_{\mu 2}&
     \frac{1}{\sqrt{2}}e^{i\delta} \cr
     \sqrt{2} U_{\mu 1} & \sqrt{2} U_{\mu 2}&
     - \frac{1}{\sqrt{2}}e^{i\delta} \cr }
\right)~~.
\end{equation}
Unitarity constraints imply that the remaining matrix elements can be
written in terms of a single mixing angle:
\begin{equation}
\label{eq:14}
  \matrix{
     U_{e 1} = \cos \theta, & U_{e 2} = - \sin \theta, \cr
     U_{\mu 1}= \sin \theta, & U_{\mu 2} = \cos \theta.  }
\end{equation}
This is the neutrino mixing matrix given in Ref. \cite{alter}.  
Setting
$\theta = \pi/4$ and $\delta = 0$ yields bi-maximal mixing of three
active neutrinos \cite{bimax1,bimax2}. With a suitably-chosen mass
hierarchy bi-maximal neutrino mixing matrix can successfully explain
observations of the atmospheric muon neutrino deficit and can rule out
the small-angle   
MSW solution of the solar neutrino data if relic neutrinos contribute
more than one percent to the closure density of the universe
\cite{georgig}. 

\vskip 1.0cm

\section{Mixing of an arbitrary number of flavors}

In this case the states
\begin{equation}
  \label{eq:22}
  \left(\matrix{
      | \nu_e \rangle \cr
      | \tilde{\nu}_{\mu} \rangle  \cr
      | \nu_s \rangle \cr
      . }
\right)~~~
\end{equation}
are obtained by transforming the mass eigenstates with the matrix 
\begin{equation}
  \label{eq:23}
 T = \left(\matrix{
     U_{e 1} & U_{e 2} &  U_{e 4} & . \cr
     U_{\mu 1}/\cos \phi & U_{\mu 2}/ \cos \phi &
      U_{\mu 4}/ \cos \phi & . \cr
       U_{s 1} & U_{s 2} &  U_{s 4} & . \cr
       . & .& .& .}
\right)~~.
\end{equation}
The constraint we have adopted, Eq. (\ref{eq:15}), is motivated by  
recent
experimental results. It is possible to put this in a form which may
motivate model building. In general N flavors of neutrinos mix under
the fundamental representation of $U(N)$. A general element of
$U(N)$ can be written as a product of $N(N-1)/2$ different,
non-commuting $SU(2)$
rotations and a diagonal matrix with matrix elements which are pure
phases. (In the familiar $U(3)$ case these SU(2) groups are known as
i-, u-, and v-spin \cite{liebook}). Thus, for $N$ flavors, one can
write 
\begin{equation}
  \label{eq:24}
  U^{\dagger}_{i\alpha} = \left(\matrix{
     C_{12} &  S_{12}^{\ast} &  0 & . \cr
     -S_{12} &  C_{12}^{\ast} & 0 & . \cr
       0 & 0&  1 & . \cr
       . & .& .& .}
\right) \left(\matrix{
     C_{13} & 0 &  S_{13}^{\ast} & . \cr
     0 &  1 & 0 & . \cr
     - S_{13} & 0&  C_{13}^{\ast} & . \cr
       . & .& .& .}\right) \cdots
 \left(\matrix{
     e^{i\delta_1} & 0 &  0 & . \cr
     0 &  e^{i\delta_2}  & 0 & . \cr
     0 & 0&  e^{i\delta_3} & . \cr
       . & .& .& .}\right) ~~.
\end{equation}
Here we label each $SU(2)$ rotation $R_{ab}, a<b, a,b=1, \cdots
N$. The $aa$-th, $bb$-th, $ab$-th, and $ba$-th elements of $R_{ab}$
are $C_{ab}, C_{ab}^{\ast}, - S_{ab}^{\ast},$ and $S_{ab}$,
respectively. In all these matrices the condition $|C_{ab}|^2 +
|S_{ab}|^2 =1$ is satisfied.  The phases $\delta_1,\delta_2,..$ may be
absorbed into the mass eigenstates. Our constraint is equivalent to
choosing $C_{a 3} =1$ for all $a$ except for $a=2$. For this case we
have $C_{2 3} = \cos \phi$ and  $S_{2 3} = e^{i \eta} \sin \phi$. This
reduces the original $N(N-1)/2$ parameters into $(N^2-3N+4)/2$
parameters.

Either using Eq. (\ref{eq:24}) or the conditions derived in the
Introduction one can easily show that the matrix $T$ of
Eq. (\ref{eq:23}) is also unitary. Using this result one can write
down the equation that governs the evolution of flavor states in
matter \cite{msw}. For the special case of two active ($\nu_e$ and
$\tilde{\nu}_{\mu}$) and one sterile ($\nu_s$) flavor states (cf. 
Eq. (\ref{eq:22})) one gets 
\begin{eqnarray}
  \label{eq:25}
  &i& \frac{\partial}{\partial x}~ \left(\matrix{
      | \nu_e \rangle \cr
      | \tilde{\nu}_{\mu} \rangle  \cr
      | \nu_s \rangle } 
\right)~~~ =  
\nonumber \\ && \left(\matrix{
     \Delta_2 |U_{e 2}|^2 + \Delta_4 |U_{e 4}|^2 + V_e & \frac{1}{\cos
       \phi} [\Delta_2 U_{e 2} U_{\mu 2}^* + \Delta_4 U_{e 4} U_{\mu
       4}^*] 
 & \Delta_2 U_{e 2} U_{s 2}^* + \Delta_4 U_{e 4} U_{s 4}^*  \cr
     \frac{1}{\cos \phi} [\Delta_2 U_{e 2}^* U_{\mu 2} + \Delta_4 U_{e
       4}^* U_{\mu 4}] & \frac{1}{\cos^2 \phi} [\Delta_2 |U_{\mu 2}|^2
     + \Delta_4 |U_{\mu 4}|^2] + V_{\mu} & \frac{1}{\cos
       \phi} [\Delta_2 U_{\mu 2} U_{s 2}^* + \Delta_4 U_{\mu 4} U_{s 4}^*] 
     \cr \Delta_2 U_{e 2}^* U_{s 2} 
     + \Delta_4 U_{e 4}^* U_{s 4}    
        &    \frac{1}{\cos
       \phi} [\Delta_2 U_{\mu 2}^* U_{s 2} + \Delta_4 U_{\mu 4}^* U_{s
       4}]  &  \Delta_2 |U_{s 2}|^2 + \Delta_4 |U_{s 4}|^2 
\cr}
\right) \nonumber \\
&&~~  \times~~\left(\matrix{
      | \nu_e \rangle \cr
      | \tilde{\nu}_{\mu} \rangle  \cr
      | \nu_s \rangle }
\right)~,  
\end{eqnarray}
where 
\begin{equation}
  \label{30}
  V_e (x) = \sqrt{2} G_F \left[ N_e (x) - N_n(x) / 2 \right], 
\end{equation}
\begin{equation}
  \label{31}
  V_{\mu} (x) =  - \frac{1}{\sqrt{2}} G_F N_n(x), 
\end{equation}
and
\begin{equation}
  \label{32}
  \Delta_i = \left( m_i^2 - m_1^2 \right) / 2 E. 
\end{equation}
(Note that we defined $ \Delta_1$ to be zero). In the next section we
comment on a particular application of this evolution equation. 

\section{Applications and Conclusions}

The utility of our constraints and their role in calculating matter
enhancement effects is best illustrated with an example from
astrophysics. Ref.   \cite{cfq} presents a $4\times 4$ neutrino mass
and mixing scheme to solve the   neutron deficit problem associated
with $r$-process nucleosynthesis from neutrino-heated supernova
ejecta. This neutrino mass scheme has a   maximally mixed, or near
maximally mixed doublet of $\nu_\mu$ and $\nu_\tau$   neutrinos split
from a lower mass doublet consisting of $\nu_e$ and a sterile species,
$\nu_s$. In this model, neutrinos emitted from the surface of a
neutron star propagate through two resonances in sequence: (1) a
$\nu_{\mu,\tau} \rightleftharpoons \nu_s$ resonance; and (2) a
$\nu_{\mu,\tau} \rightleftharpoons \nu_e$ resonance.

Because of matter effects, and because of the nature of the density
gradient in supernovae \cite{cfq}, we can approximate each resonance
as a   $3\times 3$ mixing problem. At the first resonance we envision
that the $\nu_\mu$   and $\nu_\tau$ are coupled to the mass basis
similarly, and each of these   states is coupled with the $\nu_s$ by a
mixing matrix of the form in Eq.\   (\ref{eq:13}).  Likewise, the
neutrino states at the second resonance are similarly   coupled (up to
phases) in the $3\times 3$ sector, but now with the $\nu_e$
substituted for the $\nu_s$.

The relevant $3\times 3$ sector of the overall mixing matrix at each  
resonance 
can be put in the form of Eq.\ (\ref{eq:13}) in the following manner.  
We take
$\vert \nu_3\rangle$ and $\vert \nu_4\rangle$ as the mass eigenstates  
to which
$\nu_\tau$ and $\nu_\mu$ are similarly coupled in each case. The  
other flavor
state, $\vert \nu_s\rangle$ or $\vert\nu_e\rangle$, is in addition  
coupled to
mass eigenstate $\vert \nu_i\rangle$, with $i=1$ or $i=2$,  
respectively. At
each of the resonances we can write,

\begin{equation}
\label{eqn:ckm}
U=\pmatrix{
1 & 0 & 0 \cr
0 & \cos\psi & \sin\psi \cr
0 & -\sin\psi & \cos\psi \cr
} \pmatrix{
\cos\phi & 0 & e^{i\delta}\sin\phi \cr
0 & 1 & 0 \cr
e^{-i\delta}\sin\phi & 0 & \cos\phi \cr
} \pmatrix{
\cos\omega & \sin\omega & 0 \cr
-\sin\omega & \cos\omega & 0 \cr
0 & 0 & 1 \cr
}.
\end{equation}
In the model of Ref. \cite{cfq} $\psi=\pi/4$, implying maximal mixing  
of
$\nu_\mu$ and $\nu_\tau$, and $\phi =0$, to remove CP violation  
effects. The
angle $\omega$ can be chosen to ensure adiabatic flavor amplitude  
evolution in
the supernova environment (or to fit the LSND results). We note that  
this
$3\times 3$ mixing is the same as that employed in Ref.  
\cite{georgig}, but
with the opposite order for the rotations. A mixing matrix of this  
form implies
that at each resonance we can write:
\begin{equation}
\label{eqn:sch1}
\vert \nu_\alpha\rangle=\cos\omega \vert \nu_i\rangle + \sin\omega  
\vert
\nu_3\rangle
\end{equation}
\begin{equation}
\label{eqn:sch2}
\vert \nu_\beta\rangle={{1}\over{\sqrt{2}}} {\left\{-\sin\omega \vert
\nu_i\rangle + \cos\omega \vert \nu_3\rangle +\vert  
\nu_4\rangle\right\}}
\end{equation}
\begin{equation}
\label{eqn:sch3}
\vert \nu_\gamma\rangle={{1}\over{\sqrt{2}}} {\left\{\sin\omega \vert
\nu_i\rangle - \cos\omega \vert \nu_3\rangle +\vert  
\nu_4\rangle\right\}}
\end{equation}
The relevant $3\times3$ sector at the first resonance,  
$\nu_{\mu,\tau}
\rightleftharpoons\nu_s$, has $i=1$, $\alpha =s$, $\beta=\mu$, and
$\gamma=\tau$. The second resonance, $\nu_{\mu,\tau}  
\rightleftharpoons \nu_e$,
has $i=2$, $\alpha =e$, $\beta=\mu$, and $\gamma=\tau$.

Our constraints can now be applied as above to isolate the $\vert  
\nu_4\rangle$
mass eigenstate:
\begin{equation}
\label{eqn:mixstar1}
\vert\tilde\nu_\mu\rangle\equiv {{1}\over{\sqrt{2}}} {\left\{
\vert\nu_\mu\rangle - \vert\nu_\tau\rangle\right\}} =-\sin\omega  
\vert
\nu_i\rangle + \cos\omega \vert \nu_3\rangle
\end{equation}
\begin{equation}
\label{eqn:mixstar2}
\vert \tilde\nu_\tau\rangle\equiv {{1}\over{\sqrt{2}}} {\left\{
\vert\nu_\mu\rangle + \vert\nu_\tau\rangle\right\}} =  
\vert\nu_4\rangle
\end{equation}
Let us take the first resonance as an example of how the
Mikheyev-Smirnov-Wolfenstein (MSW) \cite{msw} mechanism works in the  
\lq\lq
similar coupling\rq\rq\ case. Consider a $\nu_\mu$ neutrino emitted  
from the
neutron star surface and propagating outward. Assuming the vacuum  
mixing angle
$\omega$ is small, then in vacuum $\vert\tilde\nu_\mu\rangle$ will be  
mostly
$\vert\nu_3\rangle$.  However, at high density matter effects could  
provide a
negative contribution to the effective mass of this state, and so  
cause it to
be almost entirely the lighter mass eigenstate $\vert \nu_1\rangle$.  
Note that
the other component $\vert\tilde\nu_\tau\rangle = \vert \nu_4\rangle$  
is always
a mass eigenstate.

If the neutrino flavor amplitudes evolve adiabatically, then a  
neutrino state
which begins as a mass eigenstate will remain in this mass  
eigenstate, even if
neutrinos propagate through mass level crossings. Therefore, in our  
example,
the part of the initial $\vert \nu_\mu\rangle$ which is
$\vert\tilde\nu_\tau\rangle$ propagates directly through the  
resonance
unaltered; whereas, that part of the initial state which is
$\vert\tilde\nu_\mu\rangle$ will propagate through the first  
resonance as a
$\nu_1$, which will be mostly $\nu_s$ by the time the neutrino  
reaches the
outside of the star. Therefore, MSW resonances behave like \lq\lq
filters,\rq\rq\ passing the $\vert\tilde\nu_\tau\rangle$ state, but  
converting
the $\vert\tilde\nu_\mu\rangle$ state to a flavor different than the  
initial
flavor.
This behavior will also hold in the general case where \lq\lq similar
coupling\rq\rq\ allows us to isolate a mass eigenstate.

%%%%%%%%%%%%%%%%%%%%%%%%%%%%%%%%%%%%%%%%%%%%%%%%%%%%%%%%%%%%%%%%%
%%%%%%%%%%%%%%%%%%%%%%%%%%%%%%%%%%%%%%%%%%%%%%%%%%%%%%%%%%%%%%%%%%%

\section*{ACKNOWLEDGMENTS}

This work was supported in part by the U.S. National Science
Foundation Grants No.\ PHY-9605140 at the University of Wisconsin,  
and
PHY-9800980 at the University of California, San Diego and in part by
the University of Wisconsin Research Committee with funds granted by
the Wisconsin Alumni Research Foundation. We thank the Aspen Center
for Physics for its hospitality during the early stages of this work.
We thank Institute
for Nuclear Theory at the University of
Washington for their hospitality and Department of Energy for partial
support during the completion of this work. We also would
like to thank D. O. Caldwell, G. C. McLaughlin, Y.-Z. Qian, and P.  
Ramond  for
useful discussions.

%%%%%%%%%%%%%%%%%%%%%%%%%%%%%%%%%%%%%%%%%%%%%%%%%%%%%%%%%%%%%%%%%
%%%%%%%%%%%%%%%%%%%%%%%%%%%%%%%%%%%%%%%%%%%%%%%%%%%%%%%%%%%%%%%%%

%%%%%%%%%%%%%%%%%%%%%%%%%%%%%%%%%%%%%%%%%%%%%%%%%%%%%%%%%%%%%%%%%%
%%%%%%%%%%%%%%%%%%%%%%%%%%%%%%%%%%%%%%%%%%%%%%%%%%%%%%%%%%%%%%%%%%

\end{document}